\documentclass{article}
\usepackage{amstext,amsmath,amsthm,latexsym}
\newcommand{\CC}{CC}
\newcommand{\PCC}{PCC}
\newcommand{\TCC}{TCC}

 \newtheorem{definition}{Definition}

 \newtheorem{corollary}{Corollary}
 \newtheorem{theorem}{Theorem}
 \newtheorem{lemma}{Lemma}

\newtheorem{rem}{\sc Remark}
\newenvironment{remark}{\begin{rem}}{\end{rem}}

\newtheorem{ex}{\sc Example}

\date{}
\begin{document}


\title{Individual Communication Complexity}

\author{Harry Buhrman
\thanks{
  CWI and University of Amsterdam}
\and
Hartmut Klauck
\thanks{IAS, Princeton}
 \and
Nikolai Vereshchagin
\thanks{Moscow State University}
 \and
Paul Vit\'{a}nyi
\thanks{CWI and University of Amsterdam}
}

\maketitle

\begin{abstract}
We initiate the theory of communication complexity of individual
inputs held by the agents, rather than worst-case
or average-case. We consider total, partial, and partially correct
protocols,  one-way versus two-way,  with and without help bits.
\end{abstract}

\section{Introduction}
Assume Alice has input $x$ and Bob has input $y$ and they want
to compute a function $f(x,y)$ by communicating information
and local computation according to a fixed protocol $P = (P_A,P_B)$
where $P_A$ is the protocol executed by Alice, and $P_B$
is the protocol executed by Bob. For definiteness assume that
the requirement is that Alice outputs $f(x,y)$.
 We are only interested in minimizing
the number of bits communicated between Alice and Bob as in
\cite{Ya79,KN97}.
In the usual setting one considers the worst-case or average-case over all
inputs $x,y$ of given length $n$. However, in current
situations like replicated file systems, and cache coherence algorithms,
in multiprocessor systems and computer networks, the
worst-case or average-case are not necessarily significant.
The files or updates can be very large; but in real life
they may typically be non-random
and have considerable
regularities or correlations that allow the communicated information
to be greatly compressible.
Neither the worst-case not the
average-case may be relevant; one wants to analyze the individual case.
This gives also much more information: from the individual
case-analysis one can easily derive the worst-case and the average-case,
but not the other way around. Indeed, certain phenomena have
no counterpart in more traditional settings:  For example,
there are inputs for Bob such
that irrespective of Alice's input,
every ``simple'' total protocol requires arbitrarily  higher
communication complexity than some more ``complex'' total protocol.
Our results are expressed in terms of Kolmogorov
complexity \cite{LiVi97}, the minimal number of bits from which the data
can be decompressed by effective computation. We use the ``plain''
Kolmogorov complexity denoted as $C(x), C(x|y), C(x,y)$
for the absolute complexity of $x$, the conditional complexity of $x$ given $y$,
and the joint complexity of $x,y$. 
Increased compression of the data approximates the
Kolmogorov complexity more and more, but the actual value is uncomputable
in general.  
Given $x,y$, and assuming that Alice and Bob
have a protocol $P$ that works correctly on 
$x,y$,
we study the
{\em individual communication complexity}  $\CC^P(x,y)$
defined as the
number of bits Alice with input $x$ and Bob with input $y$
exchange using protocol $P$.
We refer to a standard definition of communication protocol \cite{KN97}.
We assume that the protocol identifies the length $n$ of the strings
on which it works. By the complexity of a protocol $P$
we mean its plain Kolmogorov complexity conditional to $n$,
denoted as $C(P|n)$.

{\bf Results and Related Work:}
We use the framework of communication complexity as in \cite{Ya79,KN97}.
As far as we are aware there is no previous work on individual communication
complexity.
We formulate a theory of individual communication
complexity, and first analyze the "mother" problem, the indentity function,
 where Alice outputs the input
of Bob. 
We look at special functions such as the inner product, random functions,
and the equality
function. 
We then turn to the question of analyzing the communication complexity, 
with respect to the best protocol of given complexity,
for the mother problem (identity function).
For total protocols that are always correct, the power of one-way protocols
equals that of two-way protocols, but for partially correct protocols
or partial protocols, two-way protocols are remarkably more powerful. 
We establish a relation with Kolmogorov's Structure function,
and the existence of strange ``non-communicable'' inputs of possibly
low Kolmogorov complexity for total
protocols---for which
the communication complexity of every total protocol
is necessarily very large (almost the literal uncompressed input
needs to be communicated) unless all
of the input is hard-wired in the protocol.
It is shown that for partial protocols
two-way is more powerful than one-way  when we use help
bits.

\section{The mother function: Identity}
We start with listing some easy facts that establish lower and upper
bounds on individual communication
complexity with respect to individual protocols $P$
expressed in terms of $C(y|n)$, $C(y|P)$ and compared to
$C(y|x)$. 
We assume that the protocols do not depend on $x,y$,
they are uniform, and they compute
the function concerned on strings of length $n$.
Let $C$ be a constant such that $C(y|n)\le n+C$ 
for all $y$.

Let $I(x,y)=y$ be the identity function: Alice with input $x$ and
Bob with input $y$ compute output $y$ by Alice.
This is the ``mother'' function:
for if Alice can compute
$I$ then she can compute every computable function $f$. 

(1) For all $n$ there is a protocol $P$ of complexity $n+O(1)$ to compute
the identity function such that for all $x,y$ of length $n$ we have
$\CC_I^P(x,y)\le C(y|n)$.\\
Indeed, assume Bob knows  $L_n=|\{p\mid |p|\le n+C,\ U(p)\text{ halts}\}|$.
($U$ is the reference universal Turing machine.) 
Then Bob can find all halting programs of length at most 
$n+C$ by enumerating them until he obtains $L_n$ halting programs.
This allows him to find a shortest program $y^*$ for $y$. 
He transmits that program to Alice and Alice computes $y$.
The complexity of this protocol is $C(L_n)+O(1)=n+O(1)$.

(2) The complexity bound $n+O(1)$  on $C(P|n)$ in item (1) is tight.
For every protocol of complexity less than $n$ the assertion of item (1) is false:
for all $P$ there are $x,y$ such that 
$\CC_I^P(x,y)\ge n$ but $C(y|P)=O(1)$ (and hence $C(y|n)\le C(P|n)+O(1)$, that 
is $C(y|n)$ is much smaller than $\CC_I^P(x,y)$  if $C(P|n)$ is much smaller than $n$).\\
Indeed, let $\epsilon$ be the empty string and let $y$ be the first string such that
$\CC_I^P(\epsilon,y)\ge n$ (by counting arguments there is such $y$). 

(3)  For every protocol $P$ to compute
identity function and for every $x,y$ we have
$\CC_I^P(x,y)\ge C(y|P)-O(1)$.\\
Let $c$ be the conversation between Alice and Bob on inputs $x,y$.
It suffices to prove that given $P,c$ we can find $y$.
It is known \cite{KN97} that the set of all pairs $(x',y')$ such that 
the conversation between Alice and Bob 
on input $(x',y')$ is equal to $c$ is a rectangle, that is, has the form $X\times Y$,
for some $X,Y\subset\{0,1\}^n$.
The set $Y$ is a one-element set, as for every  $y'\in Y$ Alice
outputs $y$ also    on the input $(x,y')$ (the output of Alice depends on $c,P,x$ only).
We can find $Y$ given $P,c$ and since $Y=\{y\}$ we are done. 

By item (2), for every protocol there are 
$x,y$ such that the right hand side of the inequality
$\CC_I^P(x,y)\ge C(y|P)-O(1)$
is much less than its left hand side, more specifically, 
$C(y|P)=O(1)$ and $\CC_I^P(x,y)\ge n$.

(4) How is $\CC_I^P(x,y)$ related to $C(y|x)$? 
By item (3) we have $\CC_I^P(x,y)\ge C(y|x)-C(P)-O(\log C(P))$ for all $x,y$.
For all $P$ this inequality is not tight for some $x,y$: there are $x,y$ 
such that $C(y|x)=O(1)$ but $\CC_I^P(x,y)\ge n$.\\
Indeed, let $x=y$. We need to prove that for some $x$ it holds
$\CC_I^P(x,x)\ge n$. For every $x$ let $c(x)$ denote the conversation
on the pair $(x,x)$.  For every $x$ the set of input pairs $(x',y')$
producing the conversation $c(x)$ is a rectangle of height 1, as we have seen in
item (3). Therefore  $c(x)$ are pairwise different for different
$x$ hence for some $x$ we have  $|c(x)|\ge n$.

(5)   However, for some $P,x,y$ the inequality 
$\CC_I^P(x,y)\ge C(y|x)-C(P|n)-O(\log C(P|n))$ is close to an equality:
for all $\alpha$ there are $P,x,y$ such that $CC_I^P(x,y)=C(y|x)-\alpha+O(1)$ and
$C(P|n)\le\alpha +O(1)$.\\
Indeed, let $x$ be some string. Let $y$ be a random string of length
$n$, independent of $x$, that is, $C(y|x)=n+O(1)$. 
Let $P$ be the following protocol:
Bob first looks whether his string $y'$ has the same prefix of length
$\alpha$ than $y$. If this is the case he sends to 
Alice 0 and then $n-\alpha$ remaining bits of $y'$ and Alice 
prefixes the $n-\alpha$ received bits by $\alpha$ first 
bits of $y$ and outputs the resulting string. 
Otherwise Bob sends to 
Alice 1 and then $y'$. The complexity of this protocol is at most
$\alpha+O(1)$, as both Alice and Bob need to know only the
first $\alpha$ bits of $y$. And we have $CC_I^P(x,y)=n-\alpha=
C(y|x)-\alpha+O(1)$. 

\section{Other functions}
Because Alice can compute every computable function once she knows  Bob's
input, we have 
$\CC^{P'}_f (x,y) \leq \CC^{P}_I(x,y)$, with $C(P') \leq C(P,f)+O(1)$.

The trivial lower bound on the individual communication complexity
of a function $f$ is $\CC^P(x,y)\ge C(f(x,y) \mid x,P)-O(1)$
(and hence $\TCC^{\alpha}_f(x,y)\ge 
C(f(x,y) \mid x)-\alpha-O(\log\alpha)$ anticipating
on a later defined notion). 
In this section we establish some nontrivial lower bounds 
on $\CC^P(x,y)$ for $P$
computing $f$ on all arguments for the inner product function, the equality function 
and for
random Boolean functions.

\subsection{Inner Product}
We extend an argument introduced in \cite{BJLV00}. 
Initially, Alice has a string $x =
x_1,\ldots,x_n$ and Bob has a string $y = y_1,\ldots,y_n$ with
$x,y \in \{0,1\}^n$. Alice and Bob
compute the inner product of $x$ and $y$ modulo 2
\[f(x,y) = \sum_{i=1}^n x_i \cdot y_i \bmod 2\]
with Alice ending up with the result. 

\begin{theorem}
Every deterministic protocol $P_f$ computing
the inner product function $f$ (without help bits)
requires at least $\CC^P (x,y) \geq C(x,y \mid P)-n-O(1)$ bits of communication
on all $x,y$.
\end{theorem}

\begin{proof}
Fix a communication protocol $P$ that computes the inner product.
Let Alice's input be  $x =
x_1\ldots x_n$ and Bob's input be $y_1\ldots y_n$. 
Run the communication protocol $P$ on $x,y$ and 
let $c(x,y)$ be the communication between Alice and Bob.
Note that $P$ can be viewed as a tree with $c(x,y)$ a path in that
tree~\cite{KN97}. Hence $c(x,y)$ form a prefix free set. 
Consider the set $S=S(x,y)$ defined by
\[
S := \{ (a,b) \mid C(a,b) = C(x,y),\ 
\text{and Alice outputs $f(x,y)$ having the conversation $C(x,y)$ and input $a$}\}.
\]
We claim that $|S|\le 2^n$.
To prove the claim assume first that $f(x,y)=0$.
Let $X$ be the first projection $X$ of $S$ and
$Y$ be the second projection of $S$.
Being an intersection of two rectangles, $S$ is a rectangle too. 
As $P$ computes $f$ we know that $f(a,b)=0$ for all $(a,b)\in S$.
In other words, every
element of $X$ is orthogonal to every element in $Y$ hence
$\text{rank}(X)+\text{rank}(Y)\le n$.
Thus $|S|=|X|\cdot|Y|\le 2^{\text{rank}(X)+\text{rank}(Y)}\le 2^n$.
Assume now that that $f(x,y)=1$.
Again $S=X\times Y$ for some $X,Y$ and  
$f(a,b)=1$ for all $(a,b)\in S$.
Subtracting $x$ from the first component of all pairs in $S$ 
we obtain a rectangle $S'$ such that  $f(a,b)=0$ for all $(a,b)\in S'$.
By above argument, we have $|S'|\le 2^n$. As $|S'|=|S|$ we are done.

Given $P$, $C(x,y)$, $f(x,y)$ and  the index of $(x,y)$ in $S$ 
we can compute $(x,y)$.
By the prefix free property,
$C(x,y)$ and  the index of $(x,y)$ can be concatenated without delimiters. 
Consequently, 
$C(x,y|P) \leq |c(x,y)|+n+O(1)$.
\end{proof}

\begin{remark}
\rm
The result of the theorem
is only significant for $C(x,y) > n$, but it cannot be improved.
Namely, $\CC^P (x,y) \geq C(x,y \mid P)-n-O(1)
= C(y \mid x,P) - (n- C(x \mid P))-O(\log n)$, where
the equality follows from the ``symmetry of information'' property of
Kolmogorov complexity \cite{LiVi97}. The term $n- C(x \mid P)$
is a called the {\em randomness deficiency} of $x$ with respect to $P$.
Clearly,
for $x=00 \ldots 0$
it is maximal with $C(x \mid P)= O(\log n)$ 
and Alice knows already from her input $x$ that $f(x,y)=0$
and no bits, or only one bit, depending on the protocol conventions,
need to be exchanged: $\CC_f^P (x,y)=0$ irrespective of the complexity of $y$
which can be $n$.
\end{remark}


\subsection{Random Functions}
Alice has $x = x_1 \ldots x_n$ and Bob has $y_1 \ldots y_n$, 
and $f: \{0,1\}^{2n} \rightarrow \{0,1\}$ satisfies 
\begin{equation}\label{eq.randomf}
C(f \mid n) \geq 2^{2n}- n.
\end{equation}
The latter condition means that the truth table describing the outcomes
of $f$ for the $2^n$ possible inputs $x$ (the rows) and the $2^n$
possible inputs for $y$ (the columns) has high Kolmogorov complexity.
If we flip the truth table for a prospective $f$ using a fair coin, then
with probability at least $1- 2^{-n}$ it will satisfy (\ref{eq.randomf}).

\begin{theorem}
Every deterministic protocol $P$ computing
a function $f$ satisfying (\ref{eq.randomf}) (without help bits) 
requires at least 
$\CC_f^P (x,y) \geq \min \{C(x \mid P), C(y \mid P)\}
- \log n-O(1)$.
\end{theorem}

\begin{proof}
Run the communication protocol $P$ on $x,y$ and 
let $c(x,y)$ be the communication between Alice and Bob.
Consider the set $S=S(x,y)$ defined by
\[
S= \{ (x',y') \mid c(x',y') = c(x,y),\ 
\text{and Alice outputs $f(x,y)$ having the conversation $c(x,y)$ and input $x'$}\}.
\]
Then $S$ is a
{\em monochromatic  rectangle} in
the function table of $f$ (that is, $f(x',y')=f(x,y)$ for all $(x',y')\in S$).
Suppose the rectangle $S$ has dimensions $a \times b$.
Then we can describe $f$ by giving $f(x,y)$, the value of $a$ in $2\log a +O(1)$
bits, the value of $b$ in $2\log b+O(1)$ bits,
the positions of the rows of the rectangles $an$ bits,
the positions of the columns of the rectangles in $bn$ 
bits, all of the table except the rectangle,
in row-major order, in $2^{2n} - ab$ bits.
This description must have length at least the
Kolmogorov complexity, so by  (\ref{eq.randomf}) we find
$$
2^{2n} - ab + (a+b)n + 2 \log ab+O(1)
\geq 2^{2n}-n.
$$
Assume w.l.o.g. that $b\ge a$.
Then $a<3n$ if $n$ is large enough.
(Otherwise we would have
$3bn\le ab\le (2b+1)n+2\log b^2+O(1)$.) 
Given the communication
sequence $c(x,y)$, $n$ and $f(x,y)$ 
we can find the rectangle $S$ that it defines. Then, we can 
reconstruct $x$ by indicating its row in the rectangle.
Then 
$C(x \mid P ) \leq |c(x,y)| + \log n+O(1)$.
\end{proof}

\subsection{Equality and Functions with Large Monochromatic Rectangles}
Let $f$ be the equality function, with $f(x,y)=1$ if $x=y$ and 0 otherwise.

\begin{theorem} 
For every deterministic protocol $P$ computing $f$ we have
$\CC^P(x,x) \geq C(x \mid P)-O(1)$ for all $x,y$.
On the other hand there is $P$ of complexity $O(1)$ such that 
there are  $x,y$ ($x \neq y$) with
$C(x \mid P), C(y \mid P) \geq n-1$ for which
$\CC_f^P(x,y) = 2$.
\end{theorem}
\begin{proof}
Lower bound: 
Since trivially the communication sequence must be different and
uniquely identify $x$ if both Alice and Bob have input $x$, we have
also $\CC_f^P(x,x) +O(1)\geq C(x \mid P)$.

Upper bound: In the function table the lower left rectangle consisting
of all $x$'s beginning with 0 and all $y$'s beginning with 1
is monochromatic (entries are all 0). Thus, a protocol where Bob communicates
one bit to Alice indicating whether $x$ starts with 0 allows Alice, in case
$y$ starts with 1, to output 0. 
Otherwise Alice and Bob start the default protocol.
Thus, for such $x,y$ and $P$ we have $\CC_f^P(x,y) = 2$.
By simple counting for some such inputs
we have $C(x \mid P),C(y \mid P) \geq n-1$.
\end{proof}

Generalizing this idea, 
every function that contains large monochromatic rectangles,
of size say $2^{2n}/n^{O(1)}$, has many pairs $x,y$ of complexity
close to $n$ for which the individual communication complexity 
drops to $O(\log n)$, as follows:

In round 1 Bob tells Alice in which large rectangle (if any) his input
is situated, by sending the index of the rectangle to Alice, and 0 otherwise.
If Bob did send an index, and Alice's input is in that rectangle as well,
then Alice outputs the color (``0'' or ``1'') of the rectangle. Otherwise,
Alice starts a default protocol.

\section{Total protocols}

Let $f$ be a function defined on pairs of strings 
of the same length. Assume that 
Alice has $x$, Bob has $y$ 
and Alice wants to compute $f(x,y)$.
As the complexity measure we consider
the number of bits communicated 
between Alice and Bob.   
The naive definition of the individual communication
complexity of the value of the function $f$ on the argument  $(x,y)$ 
is the number of communicated bits in the ``best'' communication protocol.
Then, for every $x,y$ there is a protocol with no communication
at all on $(x,y)$: the string $y$ is hard wired into the protocol.  
To meaningfully capture the individual communication 
complexity of computing a function $f(x,y)$ we define now the following
notion. 
\begin{definition}
\rm
Let $\alpha$ be a natural number parameter. Let $\TCC^{\alpha}_f(x,y)$ stand
for the minimum $\CC^{P}(x,y)$ over all total protocols
$P$ of complexity at most $\alpha$ that always
compute $f$ correctly (being total such a protocol
terminates for
all inputs, and not only for $(x,y)$).
\end{definition}

For $\alpha=n+O(1)$ we have  $\TCC^{\alpha}_f(x,y)=0$ for all computable
$f$ and all $x,y$, since we can hard wire $y$ into the protocol.
Therefore it is natural to consider only $\alpha$
that are much smaller than $n$, say $\alpha=O(\log n)$.
Since computation of the Identity function suffices to compute
all other (recursive) functions we have 
$\TCC^{\alpha+O(1)}_f (x,y) \leq \TCC^{\alpha}_I(x,y)$.
The trivial lower bound
is $\TCC^{\alpha}_f(x,y)\ge C(f(x,y) \mid x)-\alpha-O(\log\alpha)$. 
For $f=I$ this gives 
$\TCC^{\alpha}_I(x,y)\ge C(y \mid x)-\alpha-O(\log\alpha)$.

\subsection{One-way equals two-way for Identity}
Let $\TCC_{f,\text{1-way}}^\alpha(x,y)$ stand
for the minimum $\TCC^{P}(x,y)$ over all one-way (Bob sends a message to Alice)
total protocols $P$ of complexity at most $\alpha$
computing $f$ (over
all inputs, and not only on $(x,y)$).
It is clear that 
$\TCC_{f,\text{1-way}}^\alpha(x,y)$ does not depend on $x$:
indeed, consider for given $(x,y)$ the best protocol 
$P$; that protocol sends the same message 
on every other pair $(x',y)$ hence 
$\TCC^{\alpha}_{f,\text{1-way}}(x',y)\le \TCC^{\alpha}_{f,\text{1-way}}(x,y)$.
Therefore
we will use the notation $\TCC_{f}^\alpha(y)$ dropping both $x$ and ``1-way''.
Obviously, 
$$
\TCC_{f}^\alpha(x,y)\le \TCC_f^\alpha(y)
$$
for all $\alpha,x,y,f$. 

Surprisingly, for $f=I$, the Identity function,
this inequality is an equality.
That is, for total protocols ``1-way'' is as powerful as ``many-way.''
More specifically, the following holds.

\begin{theorem}\label{th1}
There is a constant $C$ such that for all $\alpha,x,y$ we have
$$
 \TCC_I^{\alpha+C}(y) \le \TCC_{I}^{\alpha}(x,y).
$$
\end{theorem}     
\begin{proof}
Pick a two-way protocol $P$ witnessing $\TCC_{I}^{\alpha}(x,y)=l$.
Let $c=c(x,y)$ be the conversation according to $P$ between Alice and Bob on inputs $x,y$.
It is known that the set of all pairs $(x',y')$ such that 
the conversation between Alice and Bob 
on input $(x',y')$ is equal to $c$ is a rectangle, that is, has the form $X\times Y$,
for some $X,Y\subset\{0,1\}^n$.
The set $Y$ is a one-element set, as for every  $y'\in Y$ Alice
outputs $y$ also    on the input $(x,y')$ (the output of Alice depends on $c,P,x$ only).

Consider the following 1-way protocol $P'$: find an $x'$ with minimum 
$c(x',y)$ and send  $c(x',y)$ to Alice. Alice 
then finds the set of all pairs $(x'',y')$ such that 
the conversation between Alice and Bob 
on input $(x'',y')$ is equal to $c(x',y)$.
As we have seen that set has the form $X\times\{y\}$ for some
$X$. Thus Alice knows $y$. As $|c(x',y)|\le|c(x,y)|=\TCC_{I}^{\alpha}(x,y)$
and $C(P'|P)=O(1)$ we are done.   
\end{proof}

\subsection{Non-Communicable objects}

The function $\TCC^{\alpha}_I(y)$, as a function of $y,\alpha$,  
essentially coincides with 
\emph{Kolmogorov structure function} $h_y(\alpha)$ 
studied in \cite{GTV01,VV02}.
The latter is defined by
$$
   h_{y}(\alpha) = \min_{S} \{\log | S| : S \ni y,\; C(S) \leq \alpha\},
$$
where $S$ is a finite set  
and $C(S)$ is 
the length (number of bits) in the 
shortest binary program from which the reference universal
machine $U$ 
computes a listing of the elements of $S$ and then
halts.
More specifically we have 
\begin{align}\label{eq.equiv}
 \TCC^{\alpha+O(1)}_I(y) &\le h_{y}(\alpha),\\
 h_{y}(\alpha+O(\log n))&\le \TCC^{\alpha}_I(y). 
\nonumber
\end{align}

To prove the first inequality
we have to transform a finite set $S\ni y$ 
into a one-way protocol $P$ of complexity at most
$\alpha = C(S)+O(1)$ witnessing
$\TCC^{\alpha}_I (y)\le\log|S|$. The protocol just 
sends the index of $y$ in $S$, or $y$ literally if $y \not\in S$.

To prove the second inequality
we have to transform
a one-way total protocol $P$ into a finite set $S\ni y$
of complexity at most $C(P)+O(\log n)$
with $\log|S|\le \CC^P(y)$.  
The set consists of all $y'$ on which $P$ sends 
the message of the same length $l$ as the length of the message on $y$.
Obviously, $|S|\le 2^{l}=2^{\CC^P(y)}$ and to specify
$S$ we need a program describing $P$ and $l$.
Thus  $C(S)\le C(P)+O(\log \TCC^{\alpha}_I(y) )\le C(P)+O(\log n)$. 
 
For the properties of $h_{y}(\alpha)$, which by Theorem~\ref{th1}
are also properties of $\TCC_{I}^\alpha(x,y)$, its relation with
Kolmogorov complexity $C(y)$ of $y$ and possible shapes 
of the function $\alpha\mapsto h_{y}(\alpha)$
we refer to~\cite{VV02}. 

We will present here only a few properties. First, two
easy inequalities:
For all $\alpha\ge O(1)$ and all $x,y$ we have 
\begin{equation}\label{eq1} 
C(y|n)-\alpha-O(\log\alpha)\le \TCC^{\alpha}_I(y)\le n-\alpha+O(1).
\end{equation}
The first inequality is the direct consequence of the 
inequality $C(y|n)\le \CC^P(y)+C(P|n)+O(\log C(P|n))$,
which is trivial.
To prove the second one consider the protocol that sends $n-\alpha+C$
bits of $y$ (for appropriate constant $C$) and the remaining
$\alpha$ bits are hardwired into the protocol.
Its complexity is at most $\alpha-C+O(1)\le\alpha$ for appropriate
choice of $C$. 

The second property is not so easy. Given $y$, consider values of $\alpha$ 
such that
\begin{equation}\label{eq.ss}
TCC^{\alpha}_I(y)+\alpha = C(y)+O(1).
\end{equation}
That is, the protocol $P$ witnessing (\ref{eq.ss}) together with 
the one-way communication
record Bob sends to Alice form a two-part code for $y$ that is---up to
an independent additive constant---as
concise as the shortest one-part code for $y$ (that has length $C(y)$
by definition). 
 Following the usage in \cite{VV02} we call
$P$ a ``sufficient'' protocol for $y$. 
The descriptions of the  protocol plus the communication precisely
describe $y$, and in fact, it can be shown that the converse holds
as well (up to a constant additive term). 
There always exists such a protocol, since the protocol that 
contains $y$ hard wired in the form of a shortest program of 
length $C(y)$ satisfies the equality with $\alpha = C(y)+O(1)$
and $TCC^{\alpha}_I(y)=0$. 
By definition we cannot have 
$TCC^{\alpha}_I(y)+\alpha < C(y)-O(1)$, but for $\alpha$ sufficiently
small we have $TCC^{\alpha}_I(y)+\alpha > C(y)+O(1)$. 
In fact, for every form of function satisfying the obvious constraints
on $TCC^{\alpha}_I$ there is a $y$ such 
that $TCC^{\alpha}_I (y)$ realizes that function up to logarithmic
precision. This shows that there are essentially non-communicable strings.
More precisely: 

\begin{theorem}\label{th11}
For every $k \leq n$ and 
monotonic decreasing function $h(\alpha)$ on
integer domain $[0,k]$ with $h(0)=n$, $h(k)=0$, $C(h)=O(\log n)$, and 
$h(\alpha)+\alpha \geq k$ for $\alpha \in [0,k]$,
there is a string $y$ of length $n$ and
$C(y)=k$  such that
\begin{align*}
 \TCC^{\alpha+O(\log n)}_I(y) &\le h(\alpha),\\
 h(\alpha+O(\log n))&\le \TCC^{\alpha}_I(y).
\end{align*}
\end{theorem}

The proof is by combining Theorem 1 of \cite{VV02} with (\ref{eq.equiv}).
In particular, for every $k < n- O(\log n)$ 
there are strings $y$ of length $n$
and complexity $k$ such that $TCC^{\alpha}_I(y) > n - \alpha$
for all $\alpha < k - O(\log n)$ while $TCC^{\alpha}_I(y) = O(1)$
for $\alpha \geq k+O(1)$. We call such strings $y$ {\em non-communicable}.
 For example,
with $k= (\log n)^2$ this shows that there are $y$ of complexity 
$C(y)= (\log n)^2$
with $TCC^{\alpha}_I(y) = n -  (\log n)^2$ for all 
$\alpha < C(y)-O(\log n)$ and $O(1)$ otherwise. 
That is, Bob can hold a highly compressible string $y$,
but cannot use that fact to reduce the communication
complexity significantly below $|y|$! Unless {\em all} information
about $y$ is hard wired in the (total) protocol the communication between
Bob and Alice requires sending $y$ almost completely literally.
For such $y$, irrespective of $x$, the communication complexity is 
{\em exponential} in the complexity of $y$ for all protocols of
complexity less that that of $y$; when the complexity of the protocol
is allowed to pass the complexity of $y$ then the communication complexity
suddenly drops to 0. 

\begin{corollary}\label{cor11}
For every $n,k$ with $k \leq n$ 
there are $y$ of length $n$ and $C(y)=k$  such that for every $x$
 $TCC_I^{\alpha} (x,y) \geq n - \alpha$ for $\alpha < C(y)-O(\log n)$; while
for every $x,y$  we have
$TCC_I^{\alpha} (x,y) = O(1)$ for $\alpha \geq C(y)+O(1)$.
\end{corollary}
This follows by combining  Theorems~\ref{th1},~\ref{th11}.
If we relax the requirement of total and correct protocols to partial and 
partially correct protocols then we obtain the significantly weaker
statements of Theorem~\ref{th61} and Corollary~\ref{co21}.

\section{Partially correct and partial protocols}

The individual communication complexity can decrease 
if we do not require the communication protocol to be correct on all the 
input pairs.
Let $\CC_{f}^{\alpha}(x,y)$ stand 
for the minimum $CC^{P}(x,y)$ over all $P$ of complexity at most $\alpha$
computing $f$ correctly on input $(x,y)$ 
(on other inputs $P$ may output incorrect result).
The minimum of the empty set is defined as $\infty$. 
Let $\CC_{f,\text{1-way}}^{\alpha}(x,y)$   
stand
for the minimum $CC^{P}(x,y)$ over all one-way (Bob sends a message to Alice)
$P$ of complexity at most $\alpha$
computing $f(x,y)$ (again, on other inputs $P$ may work incorrectly).
For instance, if $f$ is a Boolean function then 
$\CC_{f,\text{1-way}}^{O(1)}(x,y)=0$ for all $x,y$
(either the protocol outputting always 0 or  
the protocol outputting always 1 is computes $f(x,y)$ for
specific pair $(x,y)$).

\subsection{Partially correct and partial protocols versus total ones}

It is easy to see that in computing the Identity function for some $(x,y)$
total, partially correct, protocols
are more powerful than totally correct ones. 
A total partially correct protocol $P$ 
computes $f(x,y)$ correctly for some $(x,y)$,
but may err on some inputs $(u,v)$, in which case we set $\CC^P(x,y) = \infty$.
Being total such a protocol
terminates for
all inputs.

\begin{definition}
\rm
Let $\alpha$ be a natural number parameter. 
Let $\CC^{\alpha}_f(x,y)$ stand
for the minimum $\CC^{P}(x,y)$ over all total partially correct protocols
$P$ of complexity at most $\alpha$.
\end{definition}

For instance, for every $n$ there is a total protocol
$P=P_n$ computable from $n$ such 
that 
$\CC^P_{I,\text{1-way}}(x,x)=0$ 
(Alice outputs her string), thus $\CC^{O(1)}_I(x,x)=0$.
On the other hand, for random $x$ of length $n$
we have $\TCC^{\alpha}_I(x,x)\ge \TCC^{\alpha-O(1)}_I(x)\ge
C(x|n)-\alpha-O(\log\alpha)\ge n-\alpha-O(\log\alpha)$.

We also consider partial protocols that 
on some $x,y$ are allowed to get stuck, that is, give no instructions at 
all about how to proceed. Formally, such a protocol is a pair of programs
$(P_A,P_B)$. The program $P_A$ tells Alice what to do 
for each $c$ (the current part of the conversation) and 
$x$: either wait the next bit from Bob, or to send 
a specific bit to Bob, or to output a certain string and halt.
Similarly, the program $P_B$ tells Bob what to do 
for each $c$ and 
$y$: either to wait the next bit from Alice or to send 
a bit to Alice, or to halt.
This pair must satisfy the following requirements
for all $(x,y)\in\{0,1\}^n$ and all $c$:
if a party gets the instruction to send a bit
then another party gets the instruction to wait for a bit.
However we do not require that for all $(x,y)\in\{0,1\}^n$ and all $c$
both parties get some instruction, it is allowed that 
$P_A,P_B$ start some endless computation.
In particular, Alice may wait for a bit and at the same time
Bob has no instruction at all.

\begin{definition}
\rm
The complexity of a partial protocol $P=(P_A,P_B)$
is defined as $C(P|n)$.
We say that $P$ computes $f$ on the input $(x,y)$
if Alice outputs $f(x,y)$ when $P_A,P_B$ are
run on $(x,y)$. On other pairs Alice is allowed to
output a wrong answer or not output anything at all.
If protocol $P$
does not terminate, or gives an incorrect answer, for
input $(x,y)$, then $\CC^{P}(x,y)= \infty$.
Two-way and one-way individual communication complexities
with complexity of the partial protocol upper bounded by $\alpha$ 
are denoted as 
$\PCC_{f}^{\alpha}(x,y)$ 
and 
$\PCC_{f,\text{1-way}}^{\alpha}(x,y)$   
respectively.
\end{definition}
Since the total, partially correct, protocols
are a subset of the partial protocols, we always have
$\PCC_{f}^{\alpha}(x,y) \leq CC_{f}^{\alpha}(x,y) \leq \TCC_f^{\alpha}(x,y)$.
Consider again the Identity function.
We have the following obvious lower bound 
\begin{equation}\label{eq51}
C(y|x)-\alpha-O(\log\alpha)\le\PCC_{I}^{\alpha}(x,y)
\end{equation}
for all $\alpha,x,y$.
On the other hand we have the following upper
bound if $\alpha$ is at least $\log C(y)+O(1)$:
\begin{equation}\label{eq2}
\PCC_{I,\text{1-way}}^{\log C(y)+O(1)}(x,y)\le C(y).
\end{equation}
Indeed, we hardwire the value $C(y)$ in the protocol 
using $\log C(y)$ bits. This enables 
$P_B$ to find a shortest description $y^*$ of $y$
and to send it to Alice; subsequently $P_A$ decompresses the message
received from Bob. Note that the program $P_B$
gives no instruction to Bob if the complexity of Bob's
input is greater than $C(y)$. Therefore, this protocol is not total.  
Comparing Equation~\eqref{eq2} to Equation~\eqref{eq1}
we see that for $\PCC$
we have a better upper bound than for $\TCC$. It turns out 
that for some pairs $(x,y)$ the communication complexity
for totally correct (and even for partially correct) protocols
is close to the upper bound $n-\alpha$ while 
the communication complexity for
partial protocols is close to the lower bound $C(y|x)\approx \alpha\ll n$.

\begin{theorem}\label{th61}
For all $\alpha,n,x$ there are $y$ of length $n$ such that 
$\CC^{\alpha}_I(x,y)\ge n-\alpha$ and
$C(y|x)\le\alpha+O(1)$.
\end{theorem}
\begin{proof}
Fix a string $x$. By counting arguments, 
there is a string $y$ with  $\CC^{\alpha}_I(x,y)\ge n-\alpha$.
Indeed, there are less than $2^{\alpha+1}$ total protocols of
complexity at most $\alpha$. For each total protocol $P$ there are
at most $2^{n-\alpha-1}$ different $y$'s with  
$CC^P(x,y)<n-\alpha$. Therefore the total number of $y$'s with 
$\CC^{\alpha}_I(x,y)< n-\alpha$ is less than $2^{\alpha+1}2^{n-\alpha-1}=2^n$.

Let $y$ be the first string with $\CC^{\alpha}_I(x,y)\ge n-\alpha$.
To identify $y$ conditional to $x$ we only need to now the 
number of total protocols 
of complexity at most $\alpha$: given that number
we enumerate all such protocols until we find all them.
Given all those protocols and $x$ we run all of them on all pairs $(x,y)$ 
to find $\CC^{\alpha}_I(x,y)$ (here we use that the protocols are total)
for every $y$, and determine the first $y$ for which it is at least $n-\alpha$.
Hence $C(y|x)\le\alpha+O(1)$.
\end{proof} 

\begin{corollary}\label{co21}
Fix constants $C_1,C_2$ such that 
$\CC^{\log C(y)+C_1}_{I,\text{1-way}}(x,y)\le C(y)\le n+C_2$.
Applying the theorem to the empty string $\epsilon$ and to (say) 
$\alpha=2\log n$
we obtain a $y$ of length $n$ with exponential 
gap between $\CC^{2\log n}_I(\epsilon,y)\ge n-2\log n-O(1)$ and
$\PCC^{\log (n+C_2)+C_1}_{I,\text{1-way}}(\epsilon,y)\le C(y)\le\log n+O(1)$.
\end{corollary}

Using a deep result of An. Muchnik~\cite{Mu02}
we can prove that 
$\PCC^{\alpha}_{I,\text{1-way}}$ is close to $C(y|x)$
for $\alpha\ge O(\log n)$ .

\begin{theorem}[An. Muchnik]
For all $x,y$ of length $n$ there is $p$ such that 
$|p|\le C(y|x)+O(\log n)$,
$C(p|y)=O(\log n)$ and 
$C(y|p,x)=O(\log n)$, where the constants in  
$O(\log n)$ do not depend on $n,x,y$.
\end{theorem}

\begin{corollary}\label{co31}
For all $x,y$ of length $n$ we have 
$\PCC^{O(\log n)}_{I,\text{1-way}}(x,y)\le C(y|x)+O(\log n)$.
\end{corollary}
\begin{proof}
Let $p$ be the program of Muchnik's theorem,
let $q$ be the program of length $O(\log n)$ 
for the reference computer to reconstruct $p$ from 
$y$ and let $r$  the program of length $O(\log n)$ 
for the reference computer to reconstruct $y$ from the pair $(x,p)$.
The protocol is as follows: Bob finds $p$ from $y,q$
and sends $p$ to Alice; Alice reconstructs $y$ from
$x,r$. Both $q$ and $r$ are hardwired into the protocol,
so its complexity is $O(\log n)$.
This protocol is partial, as both Bob and Alice may be stuck
when reconstructing $p$ from $y',q$ and
$y$ from $x',r$.
\end{proof}

For very small values of $C(y|x),C(y)$ we can do even better using the
coloring lemma 3.9 and theorem 3.11 from \cite{BGLVZ}.

\begin{lemma}\label{lem.color}
Let $k_1,k_2$ be such that $C(x) \leq k_1$ and $C(y \mid x) \leq k_2$,
and let $m = |\{(x,y): C(x) \leq k_1, \; C(y \mid x) \leq k_2 \}|$.
For $M=2^{k_1}$, $N= 2^{k_2}$
and every $1 \leq B \leq N$
Bob can compute the recursive function
$R(k_1,k_2,m, y) = r_y \leq (N/B)e(MN)^{1/B}$ such that
Alice can reconstruct $y$ from $x, r_y,m$ and at most $b \leq \log B$
extra bits.
\end{lemma}

Using $k_1,k_2,m,y$, Bob can compute $r_y$ and
send it in $\log r_y$ bits to Alice. The latter computes $y$ from
$x,m,r_y$ using additionally $b \leq \log B$ special bits provided
by the protocol.
Then, the number of bits that need to be communicated,
1 round, from Bob to Alice, is
\[
\log r_y \leq k_2 - \log B + \frac{k_2 + k_1}{B}.
\]
The protocol $P=(P_A,P_B)$ uses $\leq 2(k_1+k_2) + b +O(1)$ bits.
\begin{corollary}
If $C(x),C(y|x) =  O(\log \log n)$ and $b= \Theta (\log \log n)$
then
$PCC^{\Theta (\log \log n)}_{I,1-way} (x,y) \leq C(y|x)- \Theta (\log \log n)$. 
\end{corollary}

\subsection{Two-way is better than one-way for partially correct protocols}

Note that for the Identity function all our upper bounds hold
for one-way protocols and all our lower bounds hold
for two-way protocols. The following question arises:
are two-way protocols more powerful than one-way ones 
(to compute the Identity function)?
Theorem~\ref{th1} implies that for total
protocol it does not matter whether the communication is one-way or two-way.
For partially correct total protocols and partial protocol
the situation is different. 
It turns out that partially correct total 
two-way protocols are stronger than even   
partial one-way protocols.

\begin{theorem}\label{th7}
For every $k,l,s$ such that
$k\ge s+l2^{s}$ there are strings $x,y$ of length
$(2^{s}+1)k$ such that 
$\CC_I^{O(1)}(x,y)\le 2^{s}\log(2k)$ but $\PCC^s_{I,\text{1-way}}(x,y)\ge l$.
\end{theorem}
\begin{proof} 
We let $x=z_0z_1\dots z_{2^{s}}$ where $z_0,\dots,z_{2^{s}}$ have length $k$
and $y=z_j00\dots0$ for some $j$. 

To prove the upper bound 
consider the following two-way protocol: 
Alice 
finds a set of indexes
$I=\{i_1,\dots,i_{2^{s}}\}$ such that for every distinct $j,m$ there is $i\in I$ 
such that $i$th bit of $z_j$ is different from $i$th bit of $z_m$ (such set
does exist, which may be shown by induction).
Then she sends to Bob the string $i_1\dots i_{2^{s}}$ and Bob sends 
to Alice $i$th bit of $y$ for all $i\in I$. Alice knows now $y$.
 
We need to find now particular $z_0,z_1,\dots, z_{2^{a+b+s}}$ such that no one-way
protocol is effective on the pair $(x,y)$ obtained from
them in the specified way. To this end let $P_1,\dots, P_N$ be
all the one-way partial protocols of complexity less than $s$ 
computing the identity function.
For every  $z$ and $i\le N$  
let 
$c(z,i)$ denote the message sent by Bob in protocol $P_i$ when he receives
$z00\dots 0$ as help bits provided the length of the message is less than $l$.
Otherwise let $c(z,i)=\infty$. Let  $c(z)$ stand for the concatenation
of  $c(z,i)$ over all  $i$.   
The range of $c(z)$ has $(2^l)^{N}<2^{l2^{s}}$ elements.
Hence there
is $c$ such that for at least $2^{k-2^{s}l}>2^{s}$ different $z$'s
we have  $c(z)=c$. Pick such $c$ and pick different
$z_0,z_1,\dots, z_{2^{s}}$ among those $z$'s. 
Let $y_j$ stand for the string obtained from $z_j$ by appending 0s.
We claim that $\CC^{P_i}_I(x,y_j)\ge l$ for some $j$ for all $i\le N$.
Assume that this is not the case. That is,
for every $j$ there are $i$ such that 
$\CC^{P_i}_I(x,y_j)<l$. 
There are $j_1\ne j_2$ for which $i$ is the same.
As $c(z_{j_1},i)=c(z_{j_2},i)\ne\infty$ 
Alice receives the same message in $P_i$ on inputs $(x,y_{j_1})$, $(x,y_{j_2})$
and should output both answers
$y_{j_1},y_{j_2}$, which is a contradiction.
\end{proof}  

\begin{corollary}
Let in the above theorem $s=(\log k)/3$ and  $l=k^{2/3}/\log k$.
These values satisfy the condition $k\ge s+l2^{s}$ and 
hence there are $x,y$ of length about $k^{4/3}$ 
with almost quadratic gap between $\CC^{O(1)}_I(x,y)\le k^{1/3}\log k$ 
and $\PCC^{(\log k)/3}_{I,\text{1-way}}(x,y)\ge k^{2/3}/\log k$.
Letting $s=\log\log k$ and  $l=k/(2\log k)$
we obtain
$x,y$ of length about $k\log k$ 
with an exponential gap between $\CC^{O(1)}_I(x,y)\le \log k\log(2k)$ 
and $\PCC^{\log k}_{I,\text{1-way}}(x,y)\ge k/(2\log k)$.
\end{corollary}

\section{Summary of some selected results for comparison}
\begin{align*}
\bullet \; \; &\forall_{x,y,\alpha} [ \TCC_I^{\alpha} (x,y) \geq
\CC_I^{\alpha} (x,y) \geq \PCC_I^{\alpha} (x,y) ] \; \text{by definition}.
\\
\bullet \; \; & \forall_{\alpha,x,y} 
[\TCC_I^{\alpha + O(1)} (y) = \TCC_I^{\alpha} (x,y) + O(1) ]\; 
\text{Theorem~\ref{th1} and discussion}.
\\
\bullet \; \; &\forall_{n,k,\alpha} \exists_{y, |y|=n, C(y)=k} \forall_{x} [ \alpha < C(y) - O(\log n)
\Rightarrow \TCC_I^{\alpha} (x,y) \geq n- \alpha ]\; 
\text{Corollary~\ref{cor11}}.
\\
\bullet \; \; &\forall_{x,y,\alpha} 
[ \alpha \geq C(y) - O(1)
\Rightarrow \TCC_{I}^{\alpha} (x,y) = O(1) ] \; \text{Corollary~\ref{cor11}}.
\\
\bullet \; \; &\forall_{n,x,\alpha} \exists_{y, |y|=n} 
[ \alpha \geq C(y|x) - O(1) \& \CC_{I}^{\alpha} (x,y) \geq n - \alpha  ]
\; \text{Theorem~\ref{th61}}.
\\
\bullet \; \; &\forall_{x,y,\alpha} 
[\PCC_{I}^{\alpha} (x,y) \geq C(y|x) - \alpha - O(\log \alpha) ] \;
\text{ (\ref{eq51})}.
\\
\bullet \; \; &\forall_{n,x,y} 
[\PCC_{I, 1-way}^{\log C(y)+O(1)} (x,y) \leq C(y) ] \;
\text{ (\ref{eq2})}.
\\
\bullet \; \; &\forall_{n} \forall_{x,y, |x|=|y|=n} 
[\PCC_{I, 1-way}^{O(\log n)} (x,y) \leq C(y|x)+O(\log n) ]
\; \text{ Corollary~\ref{co31}}.
\\
\bullet \; \; &\forall_{k,l,s: k \geq s+l2^s} \exists_{x,y, |x|=|y|= (2^s+1)k} 
[ \CC^{O(1)}_I(x,y) \leq 2^s \log (2k) \& \PCC^s_{I, 1-way} (x,y) \geq l]
\; \text{Theorem~\ref{th7}}.
\end{align*}

\section{Protocols using help bits}
A (partial) protocol with $a$ help bits for Alice $b$ help bits for Bob 
on strings of length $n$ may be defined
as  a regular (partial) protocol $P$ on inputs $u,v$ of length
$n+a,n+b$, respectively. We say that 
$P$ computes $f$ on the input $(x,y)$ if  
there are 
$h_A\in\{0,1\}^a$ and
$h_B\in\{0,1\}^b$ such according to $P$ on input $(xh_A,yh_B)$ Alice outputs $f(x,y)$.
The crucial point in this definition is that 
the help bit sequences $h_A,h_B$ may depend on the input pair $(x,y)$.
We say that $P$ computes $f$ if it computes $f$ on 
all input pairs $(x,y)\in\{0,1\}^n$.
Thus $P$ may compute many different functions. For instance there is a protocol
with 1 help bit for Alice and no help bits for Bob that 
computes every Boolean function: Alice just receives the value
of the function as the help bit and outputs it.
Define $\CC^P(x,y)$ as
the minimum of the length of conversation on input $(xh_A,yh_B)$
according to $P$ over $h_A\in\{0,1\}^a$,
$h_B\in\{0,1\}^b$ such that on input $(xh_A,yh_B)$ Alice outputs $f(x,y)$
(the minimum of the empty set is defined as $\infty$). 
We define $\TCC^{\alpha,a,b}_{f}(x,y)$ as 
the minimum $CC^{P}(x,y)$ over all $P$ of complexity at most $\alpha$
computing $f$ (over
all inputs, and not only on $(x,y)$).
Define $\TCC_{f,\text{1-way}}^{\alpha,a,b}(x,y)$
$\CC^{\alpha,a,b}_{f}(x,y)$, $\CC^{\alpha,a,b}_{f,\text{1-way}}(x,y)$
analogously (in the latter two we minimize over all $P$ of complexity at
most $\alpha$).

\subsection{Partially correct protocols versus total ones---with help bits}
In contrast to the no-help-bit
case, now the difference between totally and partially correct  
protocols in not essential:
allowing only one extra help bit 
we can effectively transform a protocol $P$ computing 
$f$ on specific input $(x,y)$
into a protocol $P'$ computing $f$ an all inputs
so that $\TCC^{P'}_f(x,y)\le CC^{P}_f(x,y)$:
at first Alice and Bob receive 1 bit of help information
(or only one of them, and in that case he/she resends
that bit to the other; in this case the right hand side of the inequality should be 
incremented by 1). If this is the case then they start $P$. Otherwise they start
the default protocol. 
So we obtain
\begin{align*}
\TCC^{\alpha+O(1),a+1,b+1}_{f}(x,y)&\le \CC^{\alpha,a,b}_{f}(x,y),\\
\TCC^{\alpha+O(1),a,b+1}_{f}(x,y)&\le \CC^{\alpha,a,b}_{f}(x,y)+1,\\
\TCC^{\alpha+O(1),a+1,b}_{f}(x,y)&\le \CC^{\alpha,a,b}_{f}(x,y)+1.
\end{align*}

The same applies to $\TCC_{\text{1-way}}$ and $\CC_{\text{1-way}}$
(except the last inequality, as now Alice is unable to send to Bob).   
Therefore we will not consider specially totally correct protocols.
We will study only values  
$\CC^{\alpha,a,b}_{f}(x,y)$, $\CC^{\alpha,a,b}_{f,\text{1-way}}(x,y)$.

Moreover, 
we can decrease $a$ by $a'$ at the expense of increasing
$\alpha$ by $a'+b'+O(\log b')$ (help bits are appending
to the program specifying the protocol), and similarly for $b$: 
\begin{align*}
\CC^{\alpha+a'+b'+O(\log a'b'),a,b}_{f}(x,y)&\le \CC^{\alpha,a+a',b+b'}_{f}(x,y),
\end{align*}
but not vice verse. The same is true for 1-way protocols.

\subsection{Partial protocols with help bits}
For partial protocol we can even decrease $\alpha$ 
at the expense of increasing
both $a$ and $b$: indeed let $p$ be the shortest
program for $(P_A,P_B)$ and let $q$ we the prefix of $p$ and 
$r$ be the remaining bits of $p$. Consider now the following 
programs $P_A',P_B'$; both receive $r$ as help an both
have $p$ hard wired.
$P_A$ appends computes $p=qr$ and decompresses $p$
and then executes $P_A$. The program $P_B$ acts in a similar way.
Note that $C(P_A',P_B')\le|q|+O(1)$.
Thus we obtain
\begin{align*}
\PCC^{\alpha+O(1),a+\alpha'+O(\log\alpha'),b+\alpha'+O(\log\alpha')}_{f}(x,y)&
\le \PCC^{\alpha+\alpha',a,b}_{f}(x,y).
\end{align*}

\subsection{Two-way is better than one-way with help bits}

\begin{theorem}
For every $k,l,s,a,b$ such that
$k\ge a+b+s+l2^{s+b}$ there are strings $x,y$ of length
$(2^{a+b+s}+1)k$ such that there is a two-way protocol
of complexity $O(1)$ with $1$ help bit (either for Alice or for Bob)
such that $\CC^P(x,y)\le 2^{a+b+s}\log(2k)+1$ but for every
one-way protocol $P$ of complexity less than $s$ 
with $a$ help bits for Alice and $b$ help bits for 
Bob we have $\CC^P(x,y)\ge l$.
\end{theorem}
\begin{proof} 
The proof is similar to the proof of the previous Theorem~\ref{th7}.

We let $x=z_0z_1\dots z_{2^{a+b+s}}$ where $z_0,\dots,z_{2^{a+b+s}}$ have length $k$
and $y=z_j00\dots0$ for some $j$. 

To prove the upper bound 
consider the following two-way protocol: if $x,y$ has not the above form
Alice receives 0 as the help bit and starts the default protocol.
Otherwise she receives 1 as the help bit and finds a set of indexes
$I=\{i_1,\dots,i_{2^{a+b+s}}\}$ such that for every distinct $j,m$ there is $i\in I$ 
such that $i$th bit of $z_j$ is different from $i$th bit of $z_m$.
Then she sends to Bob the string $1i_1\dots i_{2^{a+b+s}}$ and Bob sends 
to Alice $i$th bit of $y$ for all $i\in I$. Alice knows now $y$.

We need to find now particular $z_0,z_1,\dots, z_{2^{a+b+s}}$ such that no one-way
protocol is effective on the pair $(x,y)$ obtained from
them in the specified way. To this end let $P_1,\dots, P_N$ be
all the one-way protocols of complexity less than $s$ 
with $a$ help bits for Alice and $b$ help bits for Bob computing the identity function.
For every  $z$, $i\le N$ and $h_B$ where $h_B$ is a binary sequence
of length $b$ 
let 
$c(z,i,h_B)$ denote the message sent by Bob in protocol $P_i$ when he receives
$z00\dots 0$ as the input and
$h_B$ as help bits provided the length of the message is less than $l$.
Otherwise let $c(z,i,h_B)=\infty$. Let  $c(z)$ stand for the concatenation
of  $c(z,i,h_B)$ over all  $i,h_B$.   
The range of $c(z)$ has $(2^l)^{N2^b}<2^{l2^{s+b}}$ elements.
Hence there
is $c$ such that for at least $2^{k-2^{s+b}l}>2^{a+b+s}$ different $z$'s
we have  $c(z)=c$. Pick such $c$ and pick different
$z_0,z_1,\dots, z_{2^{a+b+s}}$ among those $z$'s. 
Let $y_j$ stand for the string obtained from $z_j$ by appending 0s.
We claim that $\TCC^{P_i}(x,y_j)\ge l$ for some $j$ for all $i\le N$.
Assume that this is not the case. That is,
for every $j$ there are $i,h_A,h_B$ such that 
$\TCC^{P_i}(x,y_j)<l$ with help bit sequences $h_A,h_B$.
There are $j_1\ne j_2$ for which the triples $(i,h_A,h_B)$
coincide. As $c(z_{j_1},i,h_B)=c(z_{j_2},i,h_B)\ne\infty$ 
Alice receives the same message in $P_i$ on inputs $(x,y_{j_1})$, $(x,y_{j_2})$,
with the help bit sequences $h_A,h_B$ and should output both answers
$y_{j_1},y_{j_2}$, which is a contradiction.
\end{proof}  

\begin{corollary}
Let in the above theorem $a=b=s=(\log k)/6$, $l=k^{1/2}/\log k$.
These values satisfy the condition $k\ge a+b+s+l2^{s+b}$ and 
hence there are $x,y$ of length about $k^{1.5}$ for which
there is a two-way protocol of complexity $O(1)$ with only one help bit
with $\CC^P(x,y)\le k^{1/3}\log k$ but 
there is no one-way protocol of complexity $(\log k)/6$ with $(\log k)/6$ help bits
both for Alice and Bob 
with $\CC^P(x,y)<k^{1/2}/\log k$.  
\end{corollary}

\end{document}